 \documentclass{ws-ijmpd}

\usepackage{graphicx}
\usepackage{epstopdf}
\usepackage{epsfig}
 \newcommand{\be}[1]{\begin{equation}\label{#1}}
 \newcommand{\ee}{\end{equation}}

\begin{document}
 \markboth{Shuang-Nan Zhang \& Shuxu Yi} {Common misunderstanding of Birkhoff's
theorem, light deflection, generalized Shapiro delay}

%
\catchline{}{}{}{}{}
%

\title{On a common misunderstanding of the Birkhoff theorem and light deflection calculation: generalized Shapiro delay and its possible laboratory test}

\author{Shuang-Nan Zhang \& Shuxu Yi}

\address{Key Laboratory of Particle Astrophysics\\
Institute of High Energy Physics\\
Chinese Academy of Sciences\\
Beijing, 100049, China\\
zhangsn@ihep.ac.cn}

\maketitle

\begin{history}
\received{Day Month Year} \revised{Day Month Year} \comby{Managing Editor}
\end{history}

\begin{abstract}
In Newtonian gravity (NG) it is known that the gravitational field anywhere inside a spherically symmetric
distribution of mass is determined only by the enclosed mass. This is also widely believed to be true in general
relativity (GR), and the Birkhoff theorem is often invoked to support this analogy between NG and GR. Here we
show that such an understanding of the Birkhoff theorem is incorrect and leads to erroneous calculations of
light deflection and delay time through matter. The correct metric, matching continuously to the location of an
external observer, is determined both by the enclosed mass and mass distribution outside. The effect of the
outside mass is to make the interior clock run slower, i.e., a slower speed of light for external observer. We
also discuss the relations and differences between NG and GR, in light of the results we obtained in this
Lettework. Finally we discuss the Generalized Shapiro delay, caused by the outside mass, and its possible
laboratory test.
\end{abstract}


\section{Introduction}\label{sec1}

The well-known and widely used Birkhoff theorem states that the ``vacuum" metric of a spherically symmetric
distribution of mass is static and identical to the Schwarzschild metric of the enclosed total gravitational
mass; the ``vacuum" region can be either outside all the mass, or interior to some or all mass. For example, the
metric inside a cavity enclosed by a spherically symmetric shell is always regarded as flat Minkowski spacetime;
an analogy with the Newtonian gravity (NG) is often made in this case, i.e., outside mass does not produce
gravity inside. With such an analogy, one would expect that the metric anywhere inside a spherically
distribution of mass is the Schwarzschild metric of the enclosed total gravitational mass. Indeed this is the
common practice in the community, as exemplified in the following well known monographs, handbook, textbooks and
book review chapter$^{1-11}$ (see also the Appendix for explicit quotations from these references on such
understanding). This is the common understanding of the Birkhoff theorem, that has been widely applied in many
calculations using General Relativity (GR), for example when doing gravitational lensing calculations for light
passing through galaxies or clusters of galaxies\cite{3}.

Following the well-known Oppenheimer-Snyder solution\cite{12}, we recently have calculated the exact solution of
gravitational collapse of pressure-less mass, in the Schwarzschild coordinates, i.e., for an external
observer\cite{12}. We found that the interior metric is influenced by both the enclosed mass and the
distribution of mass outside; the time coordinate or clock would be discontinuous at the boundary, if the metric
in an interior ``vacuum" region is written in the Schwarzschild form, as commonly done. This is different from
the common (mis)understanding of the Birkhoff theorem\cite{13,14}. Our result is not in conflict with the
well-known Lema\^{\i}tre-Tolman-Bondi (LTB) solution$^{12-14}$, which states that the metric anywhere inside a
spherically symmetric distribution of mass is determined only by the enclosed total mass and energy. This is
because the LTB solution describes the metric in the comoving coordinates\cite{13}.

In this work, we first derive the metric inside a static spherical thin shell around a central object, and then
calculate the light deflection and delay, around and through a hollow spherical thin shell for simplicity. Our
results demonstrate that the common (mis)understanding of the Birkhoff theorem for the interior metric leads to
non-physical and erroneous results. Only when the gravitational effect of the outside mass is also considered,
the calculated light deflection and delay are then physical, in contrast to NG. The light delay time, after the
effect of the outside mass is taken into account, is called generalized Shapiro delay here. We adopt $G=c=1$
throughout this work.

\section{The metric for a thin shell around a central object}\label{sec2}
We first find the metric for a static thick shell around a central object of $m_{\rm in}$, following the method
in\cite{12}. The thick shell is assumed to have a mass density $\rho(r)$, pressure $P(r)$, inner and outer
radius $a'$ and $a$, respectively. In the usual Schwarzschild coordinates, the line element can be written in
the following form,
\begin{equation}d{s^2} = B\left( r \right)d{t^2} - A\left( r \right)d{r^2} - {r^2}d{\Omega
^2}.\end{equation} The components of the energy-momentum tensor are thus \begin{displaymath} {T_{tt}} = \rho B,\
{T_{rr}} = 0,\ {T_{\theta \theta }} = {r^2}P,\ {\rm and}\ {T_{\varphi \varphi }} = {r^2}{\sin ^2}\theta P,\
\end{displaymath} respectively. Substitute the energy-momentum tensor and the line element into Einstein's field
equation, we get
\begin{equation}\begin{array}{l}
- \frac{{B''}}{{2A}} + \frac{{B'}}{{4A}}\left( {\frac{{A'}}{A} + \frac{{B'}}{B}} \right) - \frac{{B'}}{{rA}} =
{R_{tt}} =  - 8\pi
B\left( r \right)\left( {\frac{1}{2}\rho  + P} \right), \\
\frac{{B''}}{{2B}} - \frac{{B'}}{{4B}}\left( {\frac{{A'}}{A} + \frac{{B'}}{B}} \right) - \frac{{A'}}{{rA}} =
{R_{rr}} = 8\pi A\left(
r \right)\left( {P - \frac{1}{2}\rho } \right), \\
- 1 + \frac{r}{{2A}}\left( { - \frac{{A'}}{A} + \frac{{B'}}{B}}
\right) + \frac{1}{A} = {R_{\theta \theta }} =  - 4\pi {r^2}\rho,  \\
{R_{\varphi \varphi }} = {\sin ^2}\theta {R_{\varphi \varphi }}. \\
\end{array}
\end{equation}
Re-arrange the above, we get,
\begin{equation}\begin{array}{l}
\frac{{{R_{tt}}}}{{2B}} + \frac{{{R_{rr}}}}{{2A}} + \frac{{{R_{\theta
\theta }}}}{{{r^2}}} =  - 8\pi \rho,  \\
- \frac{{A'}}{{r{A^2}}} - \frac{1}{{{r^2}}} + \frac{1}{{A{r^2}}} =
- 8\pi \rho.  \\
\end{array}\end{equation}
It is obvious that\begin{equation}{\left( {\frac{r}{A}} \right)^{'}} = 1 - 8\pi \rho {r^2}.\end{equation}
Integrate the above from anywhere between the shell and the central object to outside the shell, we get,
\begin{equation}A\left( r \right) = {\left( {1 - \frac{{2M\left( r
\right)}}{r}} \right)^{ - 1}},\end{equation}
where $M(r)$ is the enclosed total mass. After substituting the
above into the equation containing $R_{\theta \theta }$, we get
\begin{equation}\frac{{B'}}{B} = \frac{{\rho {r^2} +
\frac{{M\left( r \right)}}{r} - 4\pi {r^2}\rho }}{{\frac{r}{2}\left( {1 - \frac{{2M\left( r \right)}}{r}}
\right)}}\equiv f(r).\end{equation} Since $M\left( r \right) = {m_{in}} + 4\pi \left( {{r^3} - a{'^3}}
\right)\bar \rho, $ where $\bar \rho$ is the average density inside the shell, we have,
\begin{equation}f\left( r \right) = \frac{{\bar \rho \left( {{r^2}
- \frac{{4\pi a{'^3}}}{r}} \right) + \frac{{{m_{in}}}}{r}}}{{1 - \frac{{2{m_{in}}}}{r} - \frac{{8\pi}}{r}\bar
\rho \left( {{r^3} - a{'^3}} \right)}}.\end{equation} Clearly $f(r)$ is finite for any value of $\bar \rho$,
even if $\bar \rho \to \infty$.

Therefore we have $\ln B_{a}-\ln B_{a'} = \int\limits_{a'}^a {f\left( r \right)dr}\to 0$, when $a\to a'$. i.e.,
$B(r)$ is continuous across a thin shell with infinite density. On the other hand, since $M(r)$ across a thin
shell must be discontinuous, $A(r)$ must be discontinuous across a thin shell. In other words, the time term of
the metric is always maintained continuous, but the space term is not. In the following we will focus on the
properties of a thin shell for simplicity.

As shown in Fig.~1, for the massive thin shell of mass $m_{\rm s}$ and radius $r_{\rm s}$ around a central
object with mass $m_{\rm in}$, the metric in the Schwarzschild coordinates in the two ``vacuum" regions (inside
or outside the shell) can be expressed as\cite{1,2,3,5}, \be{eq1} ds^2  = g_{tt}dt^2 +g_{rr}dr^2 - r^2 (d\theta
^2  + \sin ^2 \theta d\phi ^2 ), \ee where $g_{rr}=-(1-R/r)^{-1}$, $g_{tt}=-h(t)g_{rr}^{-1}$, $R=r_{0}=2m_{\rm
in}$ (when $r<r_{\rm s}$) or $R=2(m_{\rm s}+m_{\rm in})$ (when $r>r_{\rm s}$).

The common practice is to redefine the time coordinate $t$ to absorb the factor $h(t)$, i.e., let $h(t)=1$ for
the new time coordinate, so that $g_{tt}=-g_{rr}^{-1}$, i.e., the metric is Schwarzschild in all ``vacuum"
regions$^{1-11}$. This is the common understanding of the Birkhoff theorem [1-11]. In fact, such a choice makes
$g_{tt}$ discontinuous across the shell, certainly incorrect according to what we just have proved above.
Moreover, in calculating observational effects such as light deflection angle and delay to be discussed in the
next section, a continuous time coordinate or clock is required, and thus the metric inside the interior
``vacuum" region cannot be in the form of Schwarzschild metric, as we will show below.

For $r>r_{\rm s}$, i.e., the outside region, the redefined $t$ matches the clock of the observer at infinity
(flat Minkowsi spacetime), if we take $h(t)=1$. This is actually the physical significance of the Schwarzschild
coordinate and also the reason why an external observer must and can only use the Schwarzschild coordinate in
carrying out his/her observations\cite{13,14,18}. However, letting $h(t)=1$ for any interior region, regardless
of the exact properties of the shells, will lead to discontinuous time term of the metric at all interior
boundaries, i.e., the clocks are defined differently at both sides of a boundary\cite{13,14}; this is clearly
nonphysical, and also mathematically incorrect according to the above discussion. Therefore the common
(mis)understanding of the Birkhoff is problematic.

For $r<r_{\rm s}$ in the thin shell case, requiring that the metric is continuous at $r_{\rm s}$ yields \be{eq2}
h=\frac{r_{\rm s}-2(m_{\rm s}+m_{\rm in})}{r_{\rm s}-2m_{\rm in}}. \ee Note that both $m_{\rm s}$ and $m_{\rm
in}$ refer to their total gravitational masses. Clearly $h<1$ and depends on the mass and location of the shell,
i.e., the outside mass distribution does influence the metric (the clock) inside. Specifically, the presence of
the shell makes the inside clock run slower than without the shell.

Taking $m_{\rm in}=0$, i.e., for a hollow cavity, we have $g_{rr}(r<r_{\rm s})=-1$, but still have $h(r<r_{\rm
s})<1$, i.e., the inside clock still runs slower compared to the case without the shell, although the spacetime
inside the cavity is indeed flat. Once again we emphasize that forcing $h(r<r_{\rm s})=1$ leads to discontinuous
time term (clock) of the metric across the boundary of the shell.

\subsection{Light deflection and delay around and through an empty thin shell}\label{sec3}

In order to illustrate our point that forcing $h(r<r_{\rm s})=1$ inside the shell leads to nonphysical
consequences, here we calculate an ideal case of light deflection and delay around and through an empty thin
shell, for the cases of $h(r<r_{\rm s})=1$ and $h(r<r_{\rm s})<1$ (Eq.~(\ref{eq2})), respectively. Specifically,
we calculate both the light delay time ($\Delta t$) and the deviation ($\xi$) of the pass of light, with respect
to the case without the shell, i.e., light travels along a straight line with a constant speed ($v=c=1$), as
illustrated in Fig.~2. The equations of motion for a photon are given by \be{eq3}
\frac{d\phi}{dt}=-\frac{Jh}{g_{rr}r^2}, \ee and \be{eq4}
(\frac{dr}{dt})^2=(\frac{g_{rr}}{h}+\frac{J^2}{r^2})\frac{h^2}{g_{rr}^3}, \ee where $J$ is the integral
constant, i.e., the angular momentum of the photon, that is related to the impact parameter $L$ shown in Fig.~2.
Integrating both equations above, we can calculate $\xi(L)$ and $\Delta t(L)$, for both cases of $h(r<r_{\rm
s})<1$ (Eq.~\ref{eq2}) and $h(r<r_{\rm s})=1$ (the common (mis)understanding of the Birkhoff theorem) when light
passes through the thin shell; of course $h(r>r_{\rm s})=1$ for both cases.

Fig.~3 shows $\xi(L)$. For $h(r<r_{\rm s})<1$ (Eq.~(\ref{eq2})), the deviation or light deflection distance
$\xi$ initially increases (when $L$ decreases), corresponding to increasing deflection angle. At a certain
point, when $\xi + L=r_{\rm s}$, the deviation starts to decrease, because the spacetime inside the shell is
flat and no deflection occurs inside the shell. $\xi$ changes continuously throughout the full course. However,
forcing $h(r<r_{\rm s})=1$ results in a discontinuous change of $\xi$ when the light starts to pass through the
shell; even the sign of $\xi$ is changed. This means a sudden jump of the image of the object, clearly
nonphysical.

Fig.~4 shows the delay time $\Delta t(L)$ as the shell approaches the line of sight without the shell for both
cases. For $h(r<r_{\rm s})<1$ (Eq.~(\ref{eq2})), the delay time $\Delta t(L)$ increases continuously (when $L$
decreases)£¬ even after the light passes through the shell. This is understandable, because $h(r<r_{\rm s})<1$
means that the clock inside the shell runs slower, or the speed of light inside the shell is slower (for an
external observer). However, forcing $h(r<r_{\rm s})=1$ results in a discontinuous change and sharp decrease of
$\Delta t(L)$ when $\xi + L=r_{\rm s}$. This is again nonphysical.

The cause of the above ridiculous and nonphysical results are the direct consequence of the discontinuous time
term (clock) of the metric, due to the redefinition of the time coordinate by forcing $h(r<r_{\rm s})=1$.

\section{Summary and Discussion}\label{sec4}

The main results in this work are summarized as follows:

(1) The interior metric for a spherically symmetric distribution of mass in Schwarzschild coordinates is not the
standard Schwarzschild metric, even in the``vacuum" region between a shell and a central object, contrary to the
common (mis)understanding of the Birkhoff theorem exemplified in [1-11].

(2) Redefining the time coordinate (clock) in the interior can transform the metric into the Schwarzschild
metric, but leading to discontinuity of time coordinate or clock rate across the interior boundary. Therefore
such a treatment is nonphysical.

(3) The interior time term (clock) of the metric smoothly connected at the interior boundary is determined by
both the enclosed mass and the mass distribution outside. The outside mass distribution make the interior clock
run slower.

(4) Light deflection angle and delay time across a massive shell are calculated and compared with the
nonphysical results if the interior metric is assumed (mistakenly) as Schwarzschild metric.

In the following discussions we make several points in light of the above results.

\subsection{Analogy with Newtonian gravity}

In NG, it is known that for spherically symmetric distribution of mass the gravitational field anywhere is
determined only by the enclosed mass. As shown above and in many well-known textbooks, handbook, book review
chapter and monographs on general relativity [1-11], redefining the time coordinate (clock) in the interior can
transform the metric into the Schwarzschild metric determined by only the enclosed mass, for spherically
symmetric distribution of mass. Therefore an analogy is often made between GR and NG on this point [1-11].

However, the conclusion in NG is due to the exact $1/r^2$ law of gravity, whereas in GR this is only a weak
field limit. Another important difference between NG and GR is {\it time}. In NG {\it time} is absolute and
completely independent of the choice of coordinate frame. However in GR, {\it time} is just a coordinate and
can, in principle, be chosen differently in different reference frames. A convenient choice is to choose {\it
time} in such a way that the local speed of light is the same as that in vacuum and flat spacetime. This is
indeed the reason when the clock is redefined in the cavity so that the metric is a flat Minkowski spacetime
with $c=1$. However, as we have shown that such a redefinition of {\it time} leads to the discontinuity of the
{\it time} coordinate across the interior boundary, which is not an issue in NG.

It is also worth noting that in NG, the speed of light has no special role and a massless particle travels only
along a straight line. Therefore the light deflection and delay are not relevant in NG, although one can
nevertheless calculate light deflection angle by assuming that a photon has a gravitational mass (note that in
NG there is no equivalence between energy and mass, either inertial or gravitational) and travels at the speed
of light. The calculated deflection angle is known to be just half of that obtained in GR, which of course is
not due to the weak field approximation of the NG, but rather due to the different concepts on the nature of
gravity, spacetime and light in NG. In GR, light deflection and time delay are due to both the curvature of
space and slower clock (slower speed of light), but the gravitational mass of light is taken as zero.

\subsection{Generalized Shapiro delay and its possible test}

The Shapiro delay refers to the extra time light takes to travel through a gravitational field where the speed
of light is slower than that in vacuum without gravity (Note that this is different from the extra distance
light has to travel due to the curvature of space). In normal circumstances, the Shapiro delay is always
calculated when light is deflected around an object. However, as we have shown in this Letter, the light
deflection angle and delay are also influenced in a non-trivial way if light travels through a region beyond
which there is still non-negligible amount of mass. Therefore the outside mass distribution will causes
additional delay, as shown in Fig.~4. We call the total time delay caused by both interior and outside masses
the {\it Generalized Shapiro Delay}.

According to Eq.~(\ref{eq2}), it can be shown easily $h\cong1$ for all known astrophysical systems from which
the light cross time may be observed. For example $1-\sqrt{h}\approx10^{-6}-10^{-3}$ for light passing through
galaxies or clusters of galaxies, even in the most optimistic cases. It is thus inconceivable to test this
effect using astronomical observations. This also means that forcing $h=1$ in the interior region, though
conceptually incorrect, is a good enough approximation for the precision required currently, as done for all
gravitational lensing calculations for galaxies or clusters of galaxies.

However, it may be possible to test this effect on the Earth with hyper-precision experiments and measurements
in laboratories. For example, $1-\sqrt{h}\approx10^{-24}$ for a cavity with the shell mass of 10$^3$ kg and
radius of 1 m. Therefore the total extra delay time is, $\Delta t\approx 4r_{\rm s}(1-\sqrt{h})\approx10^{-15}$
second, for each round trip across the center of the cavity ($L=0$ in Fig.~4); bouncing light back and forth
across the cavity for thousands of times would produce an extra delay time on the order of several pico-seconds,
perhaps within the reach of the current and near-future experimental precisions, if the effect of the gravity of
the earth can be taken out properly. Alternatively one can compare two high precision clocks initially
synchronized, and then one is placed inside a cavity and one is left outside. After a sufficiently long waiting
time, the inside clock should be observed to lag behind the outside clock. This would make another laboratory
test of GR.

\begin{figure}
\begin{center}
\includegraphics[width=0.5\textwidth]{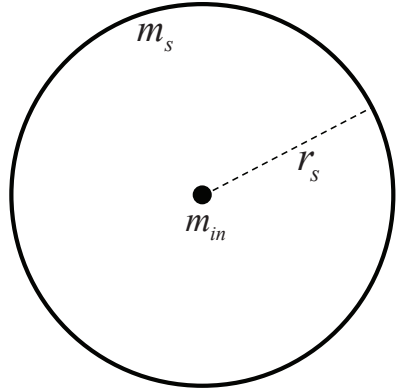}
\end{center} \caption{Illustration of a thin shell around a central object. \label{f1} }
\end{figure}

\begin{figure}
\begin{center}
\includegraphics[width=0.9\textwidth]{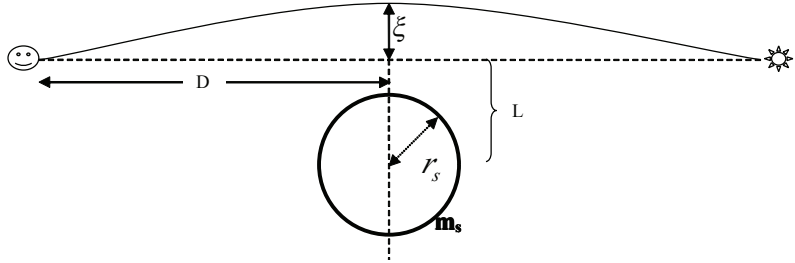}
\end{center} \caption{Illustration of light deflection around and through a thin shell. \label{f1} }
\end{figure}

\begin{figure}
\begin{center}
\includegraphics[width=0.9\textwidth]{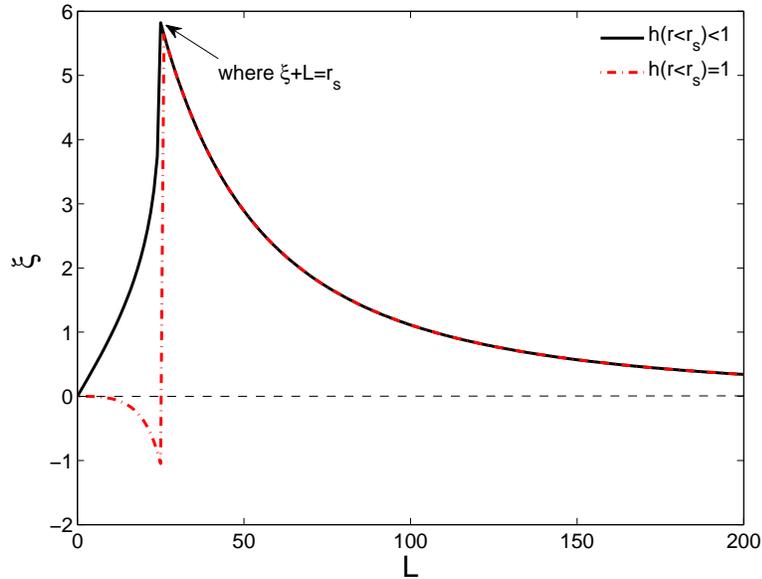}
\end{center} \caption{The light deflection distance $\xi$ (in units of $m_{\rm s}$) as a function of $L$ (in units of $m_{\rm s}$) shown in Fig.~2. $L$ is directly related to
the integral constant $J$ (angular momentum) of the photon in Eqs.~(3) and (4). The sharp turning point occurs
when $\xi + L=r_{\rm s}$ for both cases. $r_{\rm}=30 m_{\rm s}$ and $D=100 m_{\rm s}$ are assumed in the
calculation. \label{f1} }
\end{figure}

\begin{figure}
\begin{center}
\includegraphics[width=0.9\textwidth]{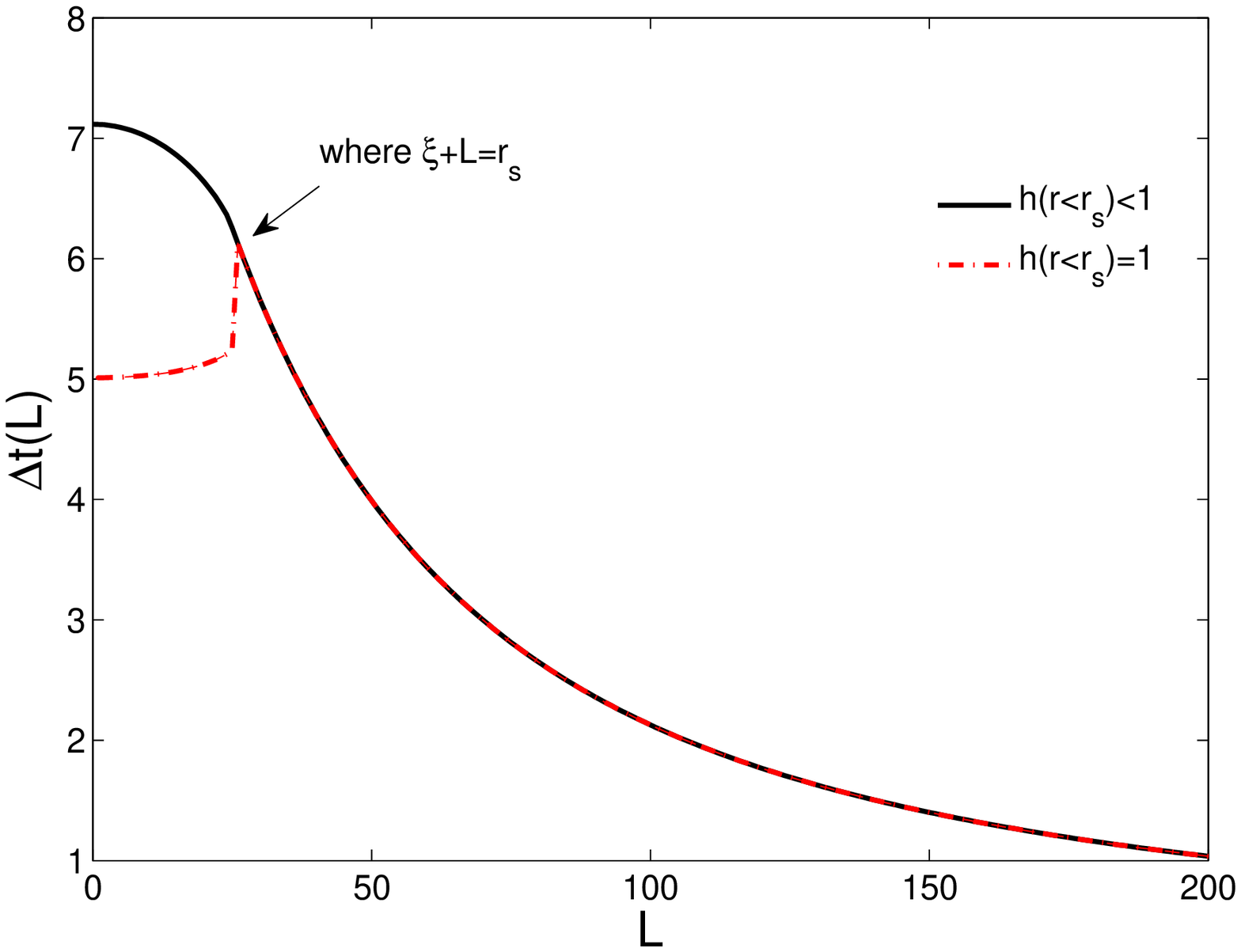}
\end{center} \caption{The light delay time $\Delta t$ (in units of $m_{\rm s}$) as a function of $L$
(in units of $m_{\rm s}$) shown in Fig.~2. $L$ is directly related to the integral constant $J$ (angular
momentum) of the photon in Eqs.~(3) and (4). The sharp turning point, for the case $h(r<r_{\rm s})=1$, occurs
when $\xi + L=r_{\rm s}$. $r_{\rm}=30 m_{\rm s}$ and $D=100 m_{\rm s}$ are assumed in the calculation.
\label{f1} }
\end{figure}

\section*{Acknowledgments}
 We thank interesting discussions with Yuan Liu, Chongming Xu and Chaoguang Huang. The anonymous referee is also thanked for
 comments that allowed us to clarify some important issues. SNZ acknowledges partial
funding support by the National Natural Science Foundation of China under project no. 11133002, 10821061,
10733010, 10725313, and by 973 Program of China under grant 2009CB824800.

\appendix

\section*{Appendix}
Here we list some quotations in references [1-11], showing that the mis-understanding of the Birkhoff Theorem is
very common and wide-spread in the community.
\begin{enumerate}
\item
H.C. Ohanian, Gravitation and Spacetime, New York \& London: W.W. NORTON \& COMPANY (1976), pp. 301:``It is a
corollary of Birkhoff's theorem that a spherically symmetric mass distribution produces no gravitational field
inside an empty spherical cavity centered on the mass distribution. This result is of course well known in the
Newtonian case... In the geometro-dynamic case, the Birkhoff theorem guarantees that the solution inside the
cavity must be of the form given by ... Since an empty cavity cannot contain any singularities, ... hence the
spacetime is flat inside the cavity."
\item
M. Harwit, Astrophysical Concepts, Fourth Edition, New York: Springer Science+Business Media, LLC,  (2006),
pp.443:``For any spherically symmetric distribution of matter in spherically symmetric motion, the dynamics
within a central sphere always remain unaffected by the distribution outside. This result, which is also valid
in general relativity and has the most wide-ranging consequences, is attributed to George Birkhoff, who first
showed its generality in what has come to be known as Birkhoff's theorem.", and pp.573:``This metric defines the
trajectories of particles and the paths along which light beams propagate in an empty Universe surrounding a
point mass. The significance of this metric, however, is far greater, as demonstrated in a powerful theorem
derived by the mathematician George D. Birkhoff in 1923. Birkhoff showed that a metric of precisely the
Schwarzschild form must hold in empty space surrounding any spherically symmetric mass distribution $M$, even
when this empty space itself is embedded in a larger, spherically symmetric distribution of matter. Moreover, he
showed that this metric must be static, invariant in time.
\item S. Weinberg,  Gravitation and Cosmology: Principles and Applications of the General Theory of
Relativity, New York: Basic Books (1977), pp.37:``According to Birkhoff's theorem, in any system that is
spherically symmetric around some point, the metric in an empty ball centered on this point must be that of flat
space. This holds whatever is happening outside the empty ball, as long as it is spherically symmetric.", and
pp.421:``According to the Birkhoff theorem, the metric and the equations of motion of a freely falling test
particle inside the sphere are independent of what is happening outside the sphere, and are therefore the same
as in a homogeneous isotropic universe, described by a Robertson-Walker metric, with a density..., and a
curvature constant that is not in general equal to the cosmological curvature constant $K$".
\item V. Mukhanov, Physical Foundations of Cosmology,
Cambridge, New York, Melbourne, Madrid, Cape Town, Singapore, Sao Paulo: Cambridge University Press (2005),
pp.9:``We assume the net effect on a particle within the sphere due to the matter outside the sphere is zero, a
premise that is ultimately justified by Birkhoff's theorem in General Relativity."
\item J.A. Peacock, An Introduction to the Physics of Cosmology, in Modern Cosmology, S. Bonometto, V.
Gorini \& U Moschella (eds), Bristol \& Philadelphia: Institute of Physics Publishing (2002), pp.22:``The
Newtonian result that the gravitational field inside a uniformshell is zero does still hold in general
relativity, and is known as Birkhoff's theorem., and pp.30:``Now look at the same situation in a completely
different way. If the particle is nearby compared with the cosmological horizon, a Newtonian analysis should be
valid: in an isotropic universe, Birkhoff's theorem assures us that we can neglect the effect of all matter at
distances greater than that of the test particle, and all that counts is the mass between the particle and us."
\item P. Coles, F. Lucchin, Cosmology: The Origin and Evolution of Cosmic Structure, Second Edition, West
Sussex, England: John Wiley \& Sons, Ltd, (2002), pp.24:``Birkhoff's theorem can also be applied to the field
inside an empty spherical cavity at the centre of a homogeneous spherical distribution of mass-energy, even if
the distribution is not static. In this case the metric inside the cavity is the Minkowski (flat-space)
metric:... This corollary of Birkhoff's theorem also has a Newtonian analogue: the gravitational field inside a
homogeneous spherical shell of matter is always zero. This corollary can also be applied if the space outside
the cavity is infinite: the only condition that must be obeyed is that the distribution of mass-energy must be
spherically symmetric."
\item C. Grupen, Astroparticle Physics, Berlin,
Heidelberg, New York: Springer (2005), pp.178:``Another non-trivial consequence of the $1/r^2$ force is that the
galaxies outside the sphere do not matter. Their total gravitational force on the test galaxy is zero. In
Newtonian gravity these properties of isotropically distributed matter inside and outside a sphere follow from
Gauss's law for a $1/r^2$ force. The corresponding law holds in general relativity as well, where it is known as
Birkhoff's theorem."
\item R. Ferraro, Einstein's Space-Time: An Introduction to
Special and General Relativity, New York: Springer Science+Business Media, LLC, (2007), pp.244:``In 1923,
Birkhoff proved that Schwarzschild solution is the only spherically symmetric vacuum solution. Therefore the
interval can also be applied inside a spherically symmetric hollow shell. But in such a case there is no reason
for the existence of a geometric singularity at the center of symmetry, what forces to choose the integration
constant M equal to zero. The space-time inside the shell has Minkowski geometry."\\
\item J.A. Peacock, Cosmological Physics, Cambridge: Cambridge University Press (1998), pp.58:``How exactly
does a black hole form once a body has become unable to support itself against its own gravity? The main
features of the problem may be understood by studying the simplest possible situation: the collapse of a star
that is taken to be a uniform pressureless sphere. The symmetry of the situation simplifies things considerably,
as does Birkhoff's theorem, which tells us that any vacuum solution of the field equations for a spherically
symmetric mass distribution is just the Schwarzschild solution, so that the field inside a spherical cavity
vanishes. The metric outside the surface of the collapsing star is thus the Schwarzschild form.", and pp.73:``In
fact, the result that the gravitational field inside a uniform shell is zero does hold in relativity, and is
known as Birkhoff's theorem."
\item P.J.E. Peebles, Physical Cosmology, Princeton: Princeton University Press (1971), pp.11:``Now imagine
that at some place there is drawn a spherical volume, radius P, and that all the matter within the sphere is
temporarily removed and set to one side. What will be the curvature of space within the evacuated sphere? The
answer is a generalization of Newton's theorem that within a hollow iron sphere the gravitational field due to
the sphere vanishes. The analogous statement in general relativity is that within a hollow centrally symmetric
system space is flat. This is a trivial application of Birkhoff's theorem."
\item P.J.E. Peebles, Principles of Physical Cosmology, Princeton: Princeton University Press (1993),
pp.63:``Birkhoff's theorem says that for a spherically symmetric distribution of matter, Einstein's field
equations have a unique solution (apart from the usual freedom of coordinate transformations). If space is empty
in some region that includes the point of symmetry, the solution in this empty hole is the flat spacetime of
special relativity.", and pp.75:``The acceleration of the radius of the sphere is given by the Newtonian
equation (D.24), because Birkhoff's theorem says the material outside the sphere cannot have any gravitational
effect on the behavior of what is inside."

\end{enumerate}
\end{document}